\documentclass[journal=jacsat,manuscript=article]{achemso}

\usepackage{chemformula} 
\usepackage[T1]{fontenc} 

\author{Patrick Maier}
\altaffiliation{Current Adress: Albert Einstein Allee 11, 89081 Ulm, Germany}
\affiliation[Ulm University]
{Institute for Quantum Optics, Ulm University}
\author{Alexander Kubanek}
\affiliation[Ulm University]
{Institute for Quantum Optics, Ulm University}

\email{patrick.maier@uni-ulm.de, alexander.kubanek@uni-ulm.de}

\title[An \textsf{achemso} demo]
  {Extracting Membrane-like hexagonal Boron Nitride hosting single Defect Centers for Resonator Integration}

\abbreviations{hBN, FPFC, AFM, NA, SPE, FIB}

\begin{document}

\begin{abstract}

The integration of membranes into optical resonators plays a key role in a variety of applications, including optomechanics. If such membranes host atom-like systems, ideally with access to spin states, new roads in quantum photonics and also in optomechanics can be taken.
Layered, two-dimensional materials have emerged as candidates for membranes hosting atom-like quantum emitters. Hexagonal boron nitride (hBN) is among the most promising two-dimensional platforms showing good mechanical properties combined with the ability to host various kinds of optical active (spin-) defects. However, the determinisitc creation of optically active defect centers in hBN membranes is an outstanding challenge. Commercially available flakes of hBN host defect centers with promising optical properties, but the integration into optical resonators suffers from scattering losses due to the flakes topography and suitable transfer, handling and manipulation techniques need to be established. Here, we develope a toolset of nano-scaled manipulation techniques to extract membrane-like structures of commercially-available hBN containing spectrally narrow single photon emitters. We demonstrate the transfer and integration into photonic devices, by coupling a single photon emitter in membran-like hBN to the mode of an open Fabry-Perot fiber cavity (FPFC) and observe cavity induced spectral enhancement by a factor of up to 100 at room temperature. Overcoming hBN-induced scattering for extracted hBN membranes, which host single photon emitters, paves the way for future applications such as its use as an optomechanical system.
\end{abstract}

\section{Introduction}

The integration of optomechanical membranes, based on silicone nitride (SiN) or graphene, into high-finesse Fabry-Perot cavities has led to advancements in the field of optomechanics \cite{Thompson2008, Karuza2012, DeAlba2016, Meyer2016, Rochau2021, Vezio2023}. Membranes have a characteristic thickness signifcantly smaller than the optical wavelength. The much larger lateral dimensions can be tailored towards specific requirements such as 
low mass or the formation of optimized mechanical modes. Recent efforts have geared towards the investigation of new kinds of materials, which not only offer particular mechanical properties, but are also capable of hosting atom-like quantum emitters. Hexagonal Boron Nitride, a 2D-material with a large bandgap, has emerged as a particularly compelling candidate due to two main reasons. \\
First, its ability to host a variety of different
optical active defect centers \citep{Cholsuk2024}. Subsets of these
show promising properties for applications like quantum sensing
\cite{Exarhos2019, Gao2022, Stern2022} and quantum communication \cite{Froech2021, Hoese2022, AlJuboori2023, Zeng2022, Samaner2022}. A variety of single photon emitters (SPEs)
\cite{Bourrellier2016,Martinez2016,Tran2016,Grosso2017,Dietrich2020,Fournier2023, Koch2024} have been investigated, some of which have shown sensitivity to applied mechanical strain \cite{Mendelson2020, Shaik2022} or optically addresable spin defects \cite{Liu2022,Gottscholl2021,Gao2022,Stern2022,Guo2023,Stern2024}. \\
Second, the hBN's promising optical as well as mechanical properties which allow the integration of hBN membranes into optomechanical systems  with advantages compared to other van der Waals materials, for example in lower photothermal heating \cite{Shandilya2019, SanchezArribas2023, Zheng2017, Falin2017, Linderaelv2021}. Creation, engineering and tuning of defect centers in hBN has been demonstrated over the last decade \cite{Bourrellier2016,Mendelson2020, Mendelson2021,Ziegler2019,Vogl2018,Chen2021,Gao2021,Kumar2023, Kianinia2020,Fournier2021,Nonahal2023,Chen2021a} and exfoliation and deterministic transfer techniques
for membranes are established \cite{MartinezJimenez2023,Fukamachi2023,Iwasaki2020,Scheuer2021a}. However, no optomechanical membrane hosting a quantum emitter in hBN, has been reported. Part of the reason is that the origin of numerous defect centers still remains unclear which makes their deterministic creation in membranes difficult. A possible solution is given by the use  of defect centers in commercially available flakes of hBN. \\
Compared to thin membranes of hBN \cite{Froech2020,Haeussler2021,Vogl2019}, the integration of these flakes into open cavities is limited by their uncontrolled topography, hindering not only the construction of optomechanical devices but also preventing exploitation of other advantages, like increased optical efficiency and spectrally narrowed lines \cite{Albrecht2013, Vogl2019,Froech2020,Haeussler2021, Kuruma2021, Froech2022}.
In this study, we demonstrate the individual extraction of membrane-like structures of hBN hosting single quantum emitters from ensembles of flakes. We transfer and couple an optically active defect center to a FPFC. We therefore deploy a toolset of manipulation techniques, leveraging a combined effort of an atomic force microscope (AFM) and a home-built manipulation system, based on a piezo driven tungsten-tip \cite{Tashima2022} with a high-NA confocal microscope. 
Coupled to the FPFC, we observe a strongly enhanced
emission rate and a reduced linewidth, in this case up to a factor of 100 and 160 respectively. These effects are attributed to cavity funneling which is enabled by the operation of the coupled system in the absence of scattering losses at the coating-defined finesse $\mathcal{F}=3400$. All these metrica are useful for developments of future optomechanical experiments with access to atomic transitions or spin states.

\section{Extraction of membrane-like hBN structures}
\label{AFM_Decluster}

Before an optically-active defect center hosted in commercially available hBN flakes can be integrated into a FPFC, individual membrane-like structures need to be prepared and transferred. Suitable defect centers are thereby pre-selected with a confocal microscope and extracted from clusters of hBN flakes with an AFM or a tungsten-tip-based manipulation system.

\begin{figure}[H]
\centering
\includegraphics[scale = 1]{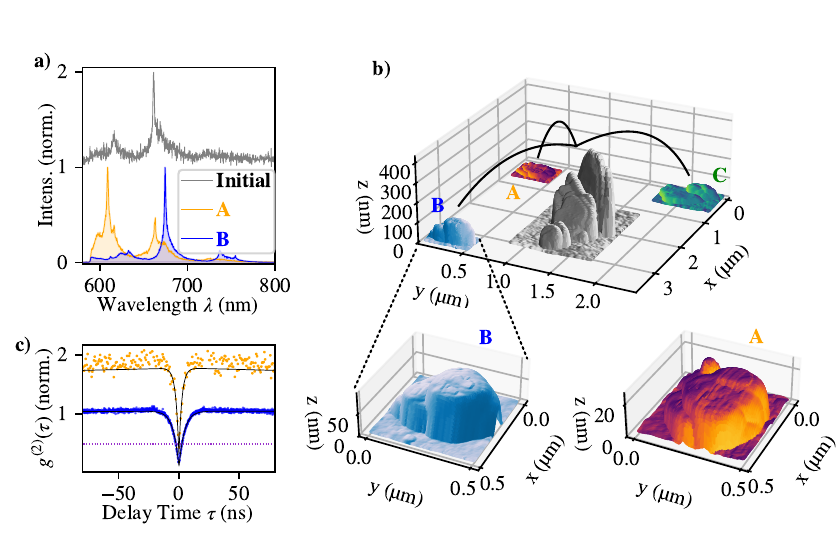}
\caption{Exemplary extraction process of membrane-like hBN structures hosting single photon emitters. \textbf{a)} Spectra of the initial ensemble of hBN particles (grey) and the extracted membrane-like structures A (orange) and B (blue). \textbf{b)} Topography of the initial ensemble (grey) and the membrane-like structures after extraction efforts with an AFM (blue, orange and green) and corresponding zoomed in visualizations of membrane-like structures A and B. \textbf{c)} Second order auto-correlation measurements for emitters A and B and corresponding fitted function, indicating single photon emission respectively.}
\label{DeClustering}
\end{figure}

As a first step a source sample of hBN flakes is prepared, where a commercially available emulsion of hBN (2D Semiconductors) is spin-coated onto a fused silica substrate after treatment in an ultrasonic bath. The spin-coated sample is annealed at 800$\,$°C for 30$\,$min under vacuum to improve the optical properties of optical active emitters \cite{Hoese2020}. As a next step the sample is investigated in a home-built confocal microscope to pre-select emitters, suitable for further integration into a FPFC according to their spectral properties. An exemplary emitter spectrum is represented in fig. \ref{DeClustering}a) (grey). To investigate the topographic properties of the pre-selected emitters host, we conduct AFM scans. As presented exemplary in fig. \ref{DeClustering}b) (grey) these scans reveal the topography of the host material, which is in this case an ensemble of a multitude of individual flakes. It is speculated that only one flake is hosting the emitter (excluding the possibility of the emitter being hosted between two or more particles). We therefore dismantle ensembles into membrane-like sub-ensembles and separate them by a few micrometers (fig. \ref{DeClustering} (blue, orange and green)) with an AFM by driving the cantilever into contact with the individual flakes. The sub-ensembles are investigated again for their optical properties (as shown exemplary for membrane-like structures A and B in fig. \ref{DeClustering} c)). A membrane-like character, with a thickness below $27\,$nm (A) and 100$\,$nm (B), thereby keeping the lateral extension in the micrometer range. In case of A, the thickness of the membrane-like structure is reduced to below a tenth of the emission wavelength of the emitter, making it particularly interesting for the integration into scattering sensitive micro-cavities. The membrane-like structures hosting emitters can then be transferred to a target sample as described in the next section. Particles without optical-active defect centers can be discarded (e.g. membrane-like structure C in fig. \ref{DeClustering} b)) and remain on the source sample.

\section{Membrane Transfer}
\label{Transfer}

\begin{figure}[H]
    \centering
    \def\svgwidth{\columnwidth}
    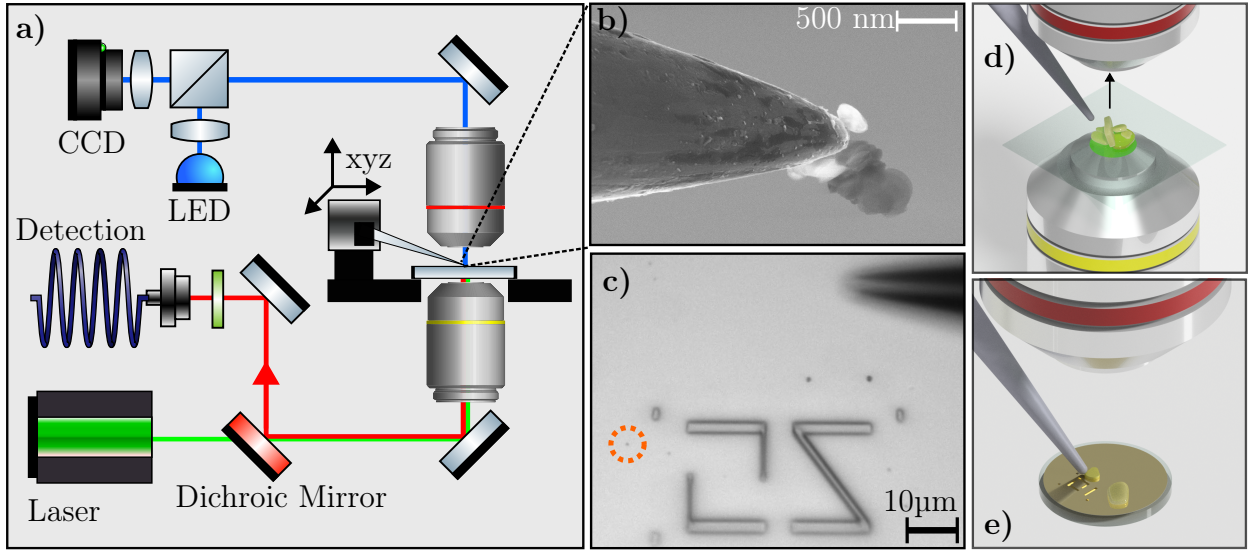
    \caption{ \textbf{a)} Schematic overview of the transfer and manipulation setup. A home-built high-NA confocal microscope is combined with a tungsten-tip mounted on a three axis piezo stage. The sample is illuminated with a $\lambda = 532\,$nm laser. An additional wide-field observation microscope is added on top of the sample to allow position control of the tungsten-tip and estimation of the size of individual hBN particles. \textbf{b)} Exemplary SEM image of a tungsten-tip with a larger ensemble of hBN flakes after lift off from a silica waver (for illustration purposes). \textbf{c)} Light microscope image of the planar cavity mirror after transfer of multiple flakes and membrane-like structures. The tungsten-tip is visible in the upper right corner. Emitter E shown in fig. \ref{Optical Table} is marked with an orange circle. \textbf{d),e):} Schematic illustration of the transfer and declustering process. With the help of the confocal microscope (yellow objective), a promising ensemble of hBN particles is localized. The tungsten-tip is geared towards the particle with the observation microscope on top (red). The tungsten-tip is brought into contact with the particle containing the previously investigated emitter, and subsequently lifted from the sample. The particle is transferred onto a macroscopic planar mirror.}
    \label{ProcessOverview}
\end{figure}

After suitable defect centers in hBN are selected, they can be integrated into optical and photonic devices (here onto a macroscopic mirror of a FPFC). The pick- and place transfer technique leverages a home-built nano-manipulation system which consists of a tungsten-tip (tip radius < 0.1$\,$µm) driven by a 3 axis piezo stage (fig. \ref{ProcessOverview}a),b) ). An additional home-built observation (wide-field) microscope on top of the sample provides optical feedback over the tip position (as illustrated in fig. \ref{ProcessOverview}a)).

The combination with a confocal microscope enables at the same time near real-time optical characterization of individual flakes and membranes. The position of the tip can be monitored by both microscopes respectively. 
The tungsten-tip allows to lift off individual clusters of hBN flakes as well as membrane-like structures from the sample and transfer them to a target substrate, here a macroscopic cavity mirror (schematically illustrated in fig. \ref{ProcessOverview} d)-e)). 
Focused ion beam (FIB) engraved marker patterns on the target mirror enable the assignability of individual emitters after a multitude of transfer processes. The compatibility of this technique with other target substrates enables the integration of hBN nanoparticles containing single photon emitters in a wide range of photonic and microscopic scaled applications \cite{Tashima2022}. Subsequent confocal microscope investigation on the target substrate ensures the continued existence of the emitter(s). With the technique described, we transfer a multitude of flakes and membranes (>30) with different spectral and topographic properties onto a macroscopic cavity mirror and investigate them with a home-built scanning cavity microscope. Similar to the separation method described in the previous section, membrane-like structures of hBN can be extracted by driving the tungsten-tip into contact and applying mechanical shifts to the ensemble. The lack of resolution compared to the AFM technique is compensated with the real-time feedback of the confocal and observation microcope. Subsequent AFM scans can then reveal the topographic features of the extracted particles if required. For visualization purpose, we de-cluster an exemplary ensemble of flakes in a scanning electron microscope (SEM) with the same tungsten-tip as used the home-built optical setup as described in the supplementary material section 1.

\section{Cavity Integration}

Our target system is an open FPFC which consists of a microscopic curved mirror on the tip of an optical single mode fiber and a macroscopic planar mirror. The conical mirror on the tip of the fiber is machined by a combination of FIB milling and CO$_2$ laser smoothing (ROC $\approx24\,$µm) \cite{Maier2025}
 and coated with a dielectric coating at Laseroptik GmbH yielding a target finesse of $\mathcal{F}_{\text{max}}\approx 3400$ (between $\lambda = 610\,$nm and $\lambda = 680\,$nm). The planar mirror is equipped with multiple membrane-like structures and flakes of hBN containing defect centers as described in the previous sections. The cavity is enclosed in a vacuum chamber combined with a home-built vibration isolation stage to minimize the impact of acoustic and thermal noise onto the system.
 
 \begin{figure}[H]
    \centering
    \def\svgwidth{\textwidth}
    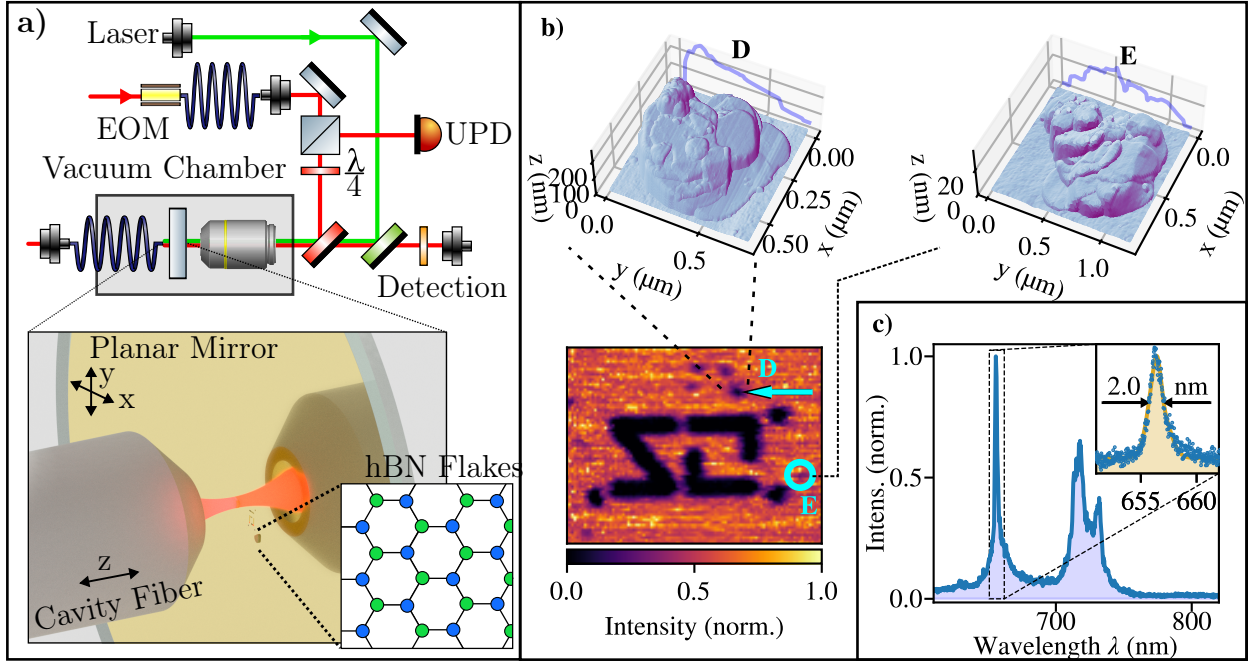
    \caption{\textbf{a)} Schematic layout of the cavity microscope. The cavity can be addressed from both sides (fiber port and planar mirror) with different lasers. An off-resonant laser with $\lambda = 532\,$nm is used for excitation. Tunable lasers are used for active stabilisation via a Pound-Drever-Hall locking scheme using an electro-optical modulator (EOM) and an ultrafast photodiode (UPD). The cavity is isolated with a vacuum chamber and a home-built vibration isolation stage to reduce the impact of surrounding mechanical and thermal noise. 
\textbf{b)} Scanning cavity microscope image of a marker pattern and adjacent hBN flakes and membrane-like structures (illuminated with a $\lambda = 609.4\,$nm laser). Flake D and membrane-like structure E are investigated in more detail with an AFM, revealing an increased thickness and a flake like structure for D compared to E, which shows a more membrane-like structure.
\textbf{c)} Optical spectrum of an emitter hosted in membrane-like structure E. \textbf{Inset:} Background corrected spectrum of the ZPL and corresponding skew Gaussian fitted function.}
\label{Optical Table}
\end{figure}

Piezo driven slick stick actuators (Attocube Systems AG) allow the precise positioning of the macroscopic planar mirror relative to the cavity mode. Scanning the planar mirror in the lateral direction of the mirror enables the system to act as a scanning cavity microscope \cite{Mader2015}. When illuminated in transmission via the fiber port, FIB engraved marker patterns as well as transferred hBN flakes and membrane-like structures can be resolved as presented in fig. \ref{Optical Table}b). Darker regions correlate to higher optical losses (mostly caused by scattering) compared to brighter regions, which is why membrane-like structures with low scattering might be not resolvable with this technique. AFM scans of flake D and membrane-like structure E confirm that an increased particle size correlates to an increased extinction rate. The indication of low optical losses induced by scattering or absorption as well as the narrow emission line ($\Delta \lambda = 2\,$nm, fig. \ref{Optical Table}c)) makes emitter E a promising candidate for coupling to the optical mode of the cavity. Once an interesting particle or membrane is selected, a piezo allows the control of the cavity length $L_{\text{Cav}}$, enabling active stabilization of the resonator with a Pound-Drever-Hall locking scheme (as schematically illustrated in fig. \ref{Optical Table} a)).

\section{Coupled System}
\label{Coupled_System}

Coupling the optical dipole of the defect center in a transferred membrane-like structure (fig. \ref{Optical Table} E) to the optical mode of the FPFC results in a narrowed and enhanced emission compared to the free-space emission. The relative position between the cavity fiber and the planar mirror are tuned with a piezo until a cavity mode and the emitter's ZPL are spectrally overlapped (fig. \ref{Coupled_System} a)). Since the emission linewidth of the coupled system is below the resolution limit of our spectrometer, the linewidth of the cavity system on the emitter is determined by locking the cavity with a laser at the next higher TEM$_{00}$ cavity resonance. The resonance on the emitter is probed with a second laser (Dye ring laser system) as presented in fig. \ref{Coupled_System} b). After fitting a Lorentzian function to the transmission data, a locked linewidth of 8.9$\,$GHz is obtained, which corresponds to a spectral narrowing factor of $\approx$160 compared to the free space emission of the emitter. 

\begin{figure}[H]
\centering
\includegraphics[scale = 1]{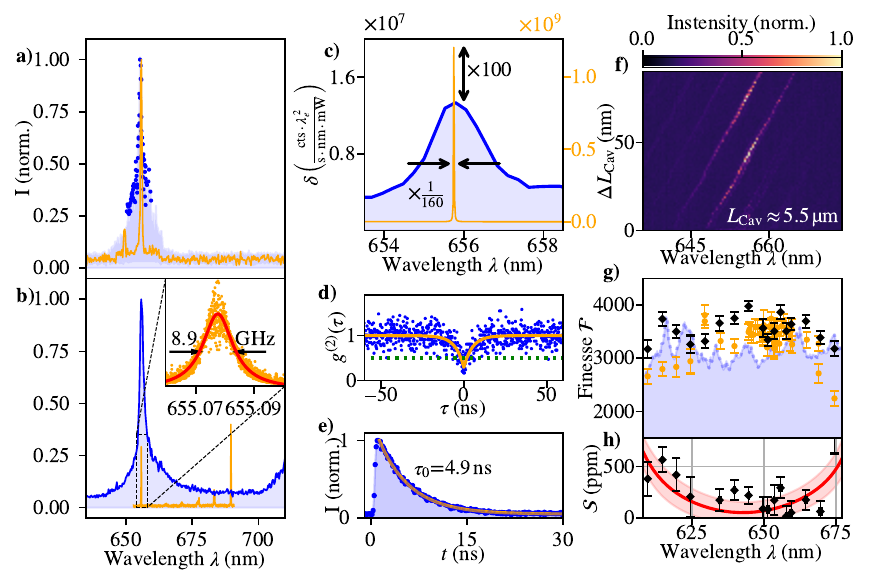}
\caption{\textbf{a)} Normalized coupled emitter-cavity spectrum (orange) of defect center E shown in fig. \ref{Optical Table} and cavity probed free space emission spectrum (blue) when exciting with a $\lambda=532\,$nm laser. \textbf{b)} Transmission spectrum when illuminating the cavity with a white light LED (orange) and free space emission spectrum of the emitter (blue). The next higher TEM$_{00}$ mode ($\lambda \approx 689\,$nm) of the cavity is used for locking. \textbf{Inset:} Spectral profile of the locked cavity on the emitter, probed with a tunable laser, revealing spectral narrowing of a factor of 160 compared to free space emission. 
\textbf{c)} Measured spectral density $\delta_{\text{sp}}$ for the emitter in free space configuration (blue) and calculated spectral density for the coupled emitter-cavity system (orange). We obtain a spectral enhancement of a factor of 100$\pm40$ induced by cavity funneling.  \textbf{d)} Normalized second order auto-correlation histogram of the coupled emitter-cavity system (blue) and corresponding fitted function (orange), which indicates single photon emission with $g^{(2)}(0)=0.3$. \textbf{e)} Pulsed lifetime measurement (start-stop histogram) of the emitter-cavity system using a pulsed laser (blue). An optical lifetime of $\tau_0 = 4.9 \pm 0.2\,$ns is extracted from a fitted exponential decay function (orange). \textbf{f)} Spectral fluorescence of the emitter cavity system for different cavity lengths $L_{\text{Cav}}$. The original cavity length $L_{\text{Cav}}\approx5.5\,$µm is tuned linearly by $\Delta L_{\text{Cav}}$. \textbf{g)} Finesse $\mathcal{F}$ of the coupled cavity-emitter system (orange) and the empty cavity (black) for different wavelengths $\lambda$. The values for the finesse are extracted from cavity transmission scans with a narrowband tunable laser. The coating-defined finesse (blue) is calculated from mirror transmission measurements performed by Laseroptik GmbH. \textbf{h)} Calculated losses $\mathcal{S}$ introduced by the integration of the emitter E (black) and corresponding fitted polynomial function (red). Losses around the ZPL of the emitter are not resolvable anymore, which would allow for experiments with even higher finesse in the future.}
\label{Coupled_System}
\end{figure}

Tuning the length of the cavity allows us to probe and reconstruct the free space emission spectrum of the emitter (depicted normalized in blue in fig. \ref{Coupled_System} a)). We compare the absolute free space emission of the emitter with the cavity coupled emission using the same optical setup and components, only flipping the mirror surface towards the objective. The spectral density $\delta$ can be calculated for both cases (free space emission and cavity coupled emission) when exciting the system off-resonantly at $\lambda = 532\,$nm, which yields a 100$\pm40\,$-fold enhanced spectral density $\delta_{\text{Cav}}$ (fig. \ref{Coupled_System} c)) in cavity configuration. This enhancement is caused by cavity funneling of the thermally broadened emission (bad emitter regime) into the cavity mode \cite{Grange2015, Janitz2020, Husel2024}.
Second-order auto correlation measurements (fig. \ref{Coupled_System} d)) reveal the single-photon emitting character of the defect center after fitting an exponential function with $g^2(0)=0.3$. Pulsed lifetime measurements indicate an optical lifetime of the coupled system of $\tau_0=4.9 \pm 0.2\,$ns which is comparable to the free space lifetime (as seen in the supplementary material section 2.). Due to the operation at room temperature the emission is thermally broadened resulting in a reduced quality factor $Q_\text{hBN}$ \cite{Albrecht2013}. This quality factor is lower compared to the cavity quality factor $Q_\text{Cav}$ (bad emitter regime).                                                                                                                                                                                                                                                                                                                                                                                                                                                                                                                                                                                                                                                                                                                                                                                                                                                                                                                                                                                                                                                                                                                                                                                                                                                                                                                                                                                                                                                                                                                                                                                                                                                                                                                                                                                                                                                                                                                                                                                                                                                                                                                                                                                                                                                                                                                                                                                                                                                                                                                                                                                                                                                                                                                                                                                                                                                                                                                                                                                                                                                                                                                                                                                   The resulting effective Purcell factor is estimated by 

\begin{equation}
F_P = \frac{3}{4\pi^2} \lambda^3 \frac{Q_{\text{eff}}}{V_m} \approx 0.6
\end{equation}

with the effective quality factor $Q_{\text{eff}}= \left(\frac{1}{Q_{\text{Cav}}} + \frac{1}{Q_{\text{hBN}}}\right)^{-1}$ and the modal volume $V_m$, which is calculated from the geometrical parameters of the cavity \cite{Auffeves2010, Huemmer2016}. To determine the impact of the host material on the system, we conduct finesse measurements and compare the coupled system with the empty cavity as a reference (fig. \ref{Coupled_System} g)). We tune the cavity length with a piezo element and record the transmitted laser intensity with an avalanche photodiode (APD). Lorentzian fits to the individual TEM$_{00}$-modes reveal the finesse by 

\begin{equation}
\mathcal{F} = \frac{\text{FSR}}{\text{FWHM}}
\end{equation}

where FSR denotes the free spectral range and FWHM the full width at half maximum of the resonant mode. The finesse is limited by the total losses of the system with 

\begin{equation}
\mathcal{F} = \frac{2\pi}{T_1 + T_2 + \mathcal{L}}
\end{equation}

where $T_1$ and $T_2$ denote the transmission coefficients of the individual mirrors. $\mathcal{L}$ describes additional losses induced e.g. by scattering and absorption. Assuming no further changes in optical losses, the losses introduced by the emitter and the host material are given by

\begin{equation}
\mathcal{L}_{\text{hBN}} = 2\pi \cdot \left( \frac{1}{\mathcal{F}}- \frac{1}{\mathcal{F}_{\text{Cav}}}\right) 
\end{equation}

where $\mathcal{F}$ denotes the measured finesse of the emitter-cavity system and $\mathcal{F}_{\text{Cav}}$ denotes the finesse of the empty cavity. Both are depicted in fig. \ref{Coupled_System}g) for different wavelengths $\lambda$ together with the coating-designed finesse given by the individual transmission of the cavity mirrors (measured by Laseroptik Gmbh). 
We directly compare finesse values for corresponding wavelengths and fit a sixth order polynomial function into each dataset (empty cavity and coupled system) to determine the optical losses for different wavelengths (depicted in fig. \ref{Coupled_System}h)). We are not able to resolve optical losses introduced by the membrane-like structure $\mathcal{L}_{\text{hBN}}$ at the ZPL wavelength of the emitter beyond the error of our measurements. The vanishing contrast in finesse between the coupled system and the empty cavity indicates that the system is no more limited in its quality factor $\mathcal{Q}$ by the introduction of the host material, which can be attributed to its membrane-like topography. \\
To further investigate the effects of the integrated material on the optical properties of the system we probe the dispersion relation of the cavity-emitter system. Linearly tuning the cavity length of the coupled system
while acquiring spectra at each step, yields a linear correlation
between the resonant frequency of the cavity-emitter
system and the length detuning (as seen in fig. \ref{Coupled_System} f)). No effect of the integration of the membrane into the resonator mode is visible in the dispersion measurement by the integration of membranes into optical resonators \cite{Haeussler2019, Koerber2023}.

\section{Conclusion}

A toolset of manipulation techniques allows us to extract membrane-like structures of hBN containing single photon emitters from clustered hBN flakes and transfer them into a FPFC. Scattering effects are reduced by manipulating the membranes topography enabling the coupling of a spectrally narrow single photon emitter in a membrane-like structure to the mode of an open FPFC. As a result, we observe spectral narrowing by a factor of 160 and spectral density enhancement by up to 100-fold as compared to free space filtering. We thereby pave the way for the integration of pre-selected quantum emitters in hBN into quantum photonic devices and also lay the foundation for the application in optomechanical experiments. Our findings are useful for future optomechanical experiments with the potential to couple the
optical dipole of the single defect-center to the motion of a suspended hBN membrane \cite{Abdi2019}. The thickness of the extracted membranes reaches below 27 nm comparable to membranes originating from mechanical exfoliation \cite{SanchezArribas2023}. The lateral dimensions reach µm-scale, about two orders of magnitude larger than the membranes thickness. However, the lateral extend remains small compared to, for example mechanically exfoliated membranes. Currently, the small lateral dimensions are about three-times smaller than the optical mode waist diameter leading to clipping of the cavity mode by the edges of the membrane, not suitable for direct implementation as optomechanical system. However, membranes with larger lateral extend could be investigated by our nanomanipulation technique. Alternatively, hybrid approaches could be investigated. The low mass in the range of femtogram together with established bonding techniques \cite{Haeussler2021, Berghaus2025} could enable hybrid optomechanical systems with an the atom-carrying system, the hBN membrane, transferred onto established optomechanical membranes, such as strained SiN. The transfer might lead to a hybridization of the modes of the high-strain SiN membrane resonator; however, the respective low mass of the hBN membrane might keep the effect small. Ultimately, such hybrid optomechanical systems could allow us to test spin-mechanical
schemes \cite{Wang2020} with the ability to engineer spin-motion interaction enabling, for example, ground-state cooling of the mechanical resonator \cite{Abdi2017}.

\begin{acknowledgement}

The authors gratefully acknowledge support of the Baden-Wuerttemberg Stiftung gGmbH in project AmbientCoherentQE. The AFM was funded by the DFG. We thank Prof. Dr. Kay Gottschalk for support. We thank Niklas Lettner, Lukas Antoniuk and Gregor Bayer for experimental support. We thank Sven Pernes, Wolfgang Rapp and the scientific workshop of the University of Ulm for technical support. Prof. Dr. Christine Kranz, Dr. Gregor Neusser and the Focused Ion Beam Center UUlm are acknowledged for their scientific support during FIB
milling. Manuel Mundszinger is acknowledged for support during SEM imaging. Measurements were conducted among others with the Qudi software suite \cite{Binder2017}. AFM scans were evaluated among others with the open source software Gwyddion \cite{Necas2012}.

\end{acknowledgement}

\begin{suppinfo}

The following files are available free of charge.
\begin{itemize}
  \item Suplemental Material: Additional resources to the main article. 

\end{itemize}

\end{suppinfo}

\bibliography{achemso-demo}

\end{document}